\documentclass{kluwer}
\usepackage{epsf}
\usepackage{rotate}

\begin{document}
\begin{article}
\begin{opening}
\title{Young close-by neutron stars:\\
the Gould Belt vs. the Galactic disc}            

\author{S.B.\surname{Popov}\email{polar@sai.msu.ru; popov@pd.infn.it}}
\institute{Universit\`a di Padova, Italy;
Sternberg Astronomical Institute, Russia}
\author{R. \surname{Turolla}} 
\institute{Universit\`a di Padova, Italy}                               
\author{M.E.\surname{Prokhorov}}
\institute{Sternberg Astronomical Institute, Russia}
\author{M. \surname{Colpi}}
\institute{Universit\`a di Milano-Bicocca, Italy}
\author{A. \surname{Treves}}
\institute{Universit\`a dell'Insubria, Italy}

\runningtitle{Young close-by neutron stars}
\runningauthor{Popov et al.}

\begin{ao}
Universit\'a di Padova\\
Dipartimento di Fisica\\
via Marzolo 8\\
35131 Padova, Italy
\end{ao}

\begin{abstract} 
 We present new population synthesis calculations of close young neutron 
stars. In comparison with our previous investigation we use a different
neutron star mass spectrum and  different initial spatial and velocity
distributions.
The results confirm that most of ROSAT dim radioquiet
isolated neutron stars had their origin in the Gould Belt.
Several tens  of young neutron stars can be identified in future
in ROSAT data at low galactic latitudes and some of them also can be
EGRET unidentified sources.
\end{abstract}


\end{opening}

\section{Introduction}

Over the last decade X-ray missions revealed an increasing number of 
isolated neutron stars (INSs) in the solar vicinity. Many of these sources,
essentially discovered by ROSAT, are not observed as active 
radio pulsars and show quite peculiar emission properties, both
at X-ray and optical wavelengths (see e.g. \opencite{t99}, \opencite{bp02},
and \opencite{h03} for recent reviews). 
Their spectrum is peaked at $\sim 100$~eV and
is well described in terms of a featureless blackbody. 
The optical emission (when observed, see \opencite{kaplan03}) 
appears close to a 
Rayleigh-Jeans distribution but lies well above the extrapolation of the
X-ray blackbody to optical wavelengths. 

The many puzzling features of X-ray emitting INSs offer contrasted
views about their nature. Although several interpretations have been 
proposed (in terms of old INSs accreting the interstellar medium,
decaying magnetars or even quark stars), the more conservative
explanation is that they are conventional middle-aged ($\approx
10^5-10^6$~yrs) cooling NSs which for reasons not understood as yet
fail to be detected as radio emitters.

A possible problem with this latter scenario is connected with the
observed overabundance of these sources in the solar proximity with respect
to what predicted by population synthesis models. If INSs are born
in the galactic disc (at about the solar distance from the galactic center) 
at the same rate at which radio pulsars are formed and if they follow 
a standard cooling history then the number of detectable X-ray sources
falls short of the observed one by about a factor a few 
(\opencite{nt99}, \opencite{p00b}). 
A possible solution is to invoke a recent epoch of
enhanced NS formation in $\lesssim 1$~kpc around the Sun. Originally this
idea has been suggested by \inlinecite{gren00} and \inlinecite{geh00}
in connection with the possibility that
unidentified EGRET sources are young close-by NSs. The Gould Belt, a
collection of young star associations which encompasses the Sun, appears
the most likely birthplace for the majority of these NSs (see the Belt
description in \opencite{p97}).
In \inlinecite{p02} it was suggested that INSs observed by ROSAT as dim
X-ray sources can be explained as young cooling objects originated
mainly from the Gould Belt.
Very recently \inlinecite{p03} (hereafter Paper I) 
addressed this issue in detail by means
of a population synthesis model in which NS formation in the Belt
(in addition to that in the galactic disc) was properly accounted for.

In this paper we present some refinements to the results of Paper I. 
In particular we explore the effects of modifying and relaxing 
some of the original assumptions contained in paper I on the computed 
Log~N~--~Log~S of cooling NSs. A central point in this respect is the
assumed NS mass spectrum since the cooling evolution is very sensitive to
the star mass. However, as it is shown, new results largely 
agree with previous ones and offer further support to the idea that
the Gould Belt is the nursery of the local NS population. 

\section{Model}

The details of our model have been presented in
Paper I; its main features are summarized below.
We assume that NSs are continuously born in the galactic disc (up 
to a distance of $\sim 3$~kpc from the Sun) and in the Gould Belt 
at a constant rate (see fig.~1). 
The rates are different in the disc and in the Belt and 
have been estimated from available SN progenitors
counts (\opencite{tam94} and 
\opencite{gren00}). Both the spatial and the cooling history
of newborn NSs is then followed as they evolve in the galactic potential.
Typically we calculate $\sim 10^4$ evolutionary tracks and then normalize 
our results to the actual number of NSs born in the considered volume 
($\sim 1000$ NSs in a sphere of radius 3 kpc centered on the Sun) 
during a 4.25 Myrs time interval. NSs cooling curves by 
\inlinecite{kam02} have 
been used to derive the NS temperature at each time step. The duration
of the calculation is fixed by
the request that the surface temperature of the lightest 
(i.e. the hottest) NSs is higher than $10^5$~K.
Cooler NSs could not have been detected by ROSAT even if they are as close as 
10 pc. Since young cooling NSs are expected to emit most of their
luminosity at UV/soft X-ray energies ($\sim 20-200$~eV,
corresponding to temperatures  $\sim 10^5$--$10^6$~K),
interstellar absorption must be accounted for. 
The PSPC count rate is finally obtained from the unabsorbed flux, which  
corresponds to the given temperature, radius and distance of the star,
and from the value of the column density. This allows us to construct
the Log~N~--~Log~S curve for close-by, cooling NSs.

Results presented in Paper I rely on  a particular choice for a set
of free parameters which enter the model. They reflect our incomplete
knowledge of some properties of the NS population, and are mainly 
related to: i) 
the spatial distribution of NS progenitors; ii) the NS mass distribution; 
iii) the NS kick velocity distribution; iv) the NS emitted spectrum. In 
Paper I we assumed that NSs are uniformly born in the Belt, modeled as a 
thin disc 500 pc in radius, and that their initial velocity distribution is 
represented by a single maxwellian with a mean velocity of 225 km~s$^{-1}$. 
The mass spectrum was taken to be flat 
in the mass range $1.1 M_\odot \leq M 
\leq 1.8 M_\odot$. Cooling NSs were assumed to emit a pure blackbody spectrum
without allowance for possible deviations arising from reprocessing in an
atmosphere and/or by a reduce surface emissivity.

\begin{figure}
\epsfxsize=\hsize
\centerline{{\epsfbox{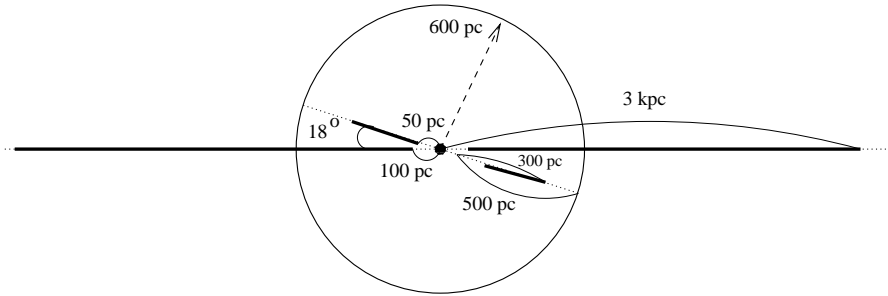}}}
\caption{A sketch of the initial spatial distribution. It is a projection
to the plane perpendicular to the galactic one. Stars are born in the Gould
Belt, which is inclined to the galactic plane by 18 degrees, and in the
galactic disc. Star producing regions are shown with thick lines.
}
\end{figure}

Here we address all these points in more detail. In particular we assess
the effects of relaxing the assumptions of Paper I on the computed 
Log~N~--~Log~S curve. The main changes are described below. As it will be
shown in the next section, the original results presented in Paper I
are not much influenced. With respect to Paper I we introduce four 
modifications:
\begin{itemize}

\item we use a slightly different spatial distribution of NS progenitors 
      taking the Gould Belt radius to be 300 pc (instead of 500 pc, see
      fig.~1). The total birthrate in the Belt was the same;

\item natal kicks were drawn from the complete velocity distribution of
      \inlinecite{acc02}, described by two maxwellians 
with total 
      average velocity $\sim 540$ km~s$^{-1}$,
      instead of a single maxwellian with average velocity $\sim 225$
      km~s$^{-1}$;

\item we account for the possible reduced emissivity of the
      star surface, as suggested by the case of RX J1856.5-3754
      (e.g. \opencite{dra02}).
      This has been mimicked using a radiation radius $R_{rad}\sim
      0.32 R$, so that $L=4\pi R_{rad}^2\sigma T_{eff}^4\sim 0.1
      L_{BB,R}$;

\item a more realistic mass spectrum of NSs, 
      peaked around $1.3$--$1.4 M_\odot$ (see fig.~2), has been derived
      and incorporated in the simulations
      instead of the flat one used before.

\end{itemize}

\begin{figure}
\epsfxsize=\hsize
\centerline{{\epsfbox{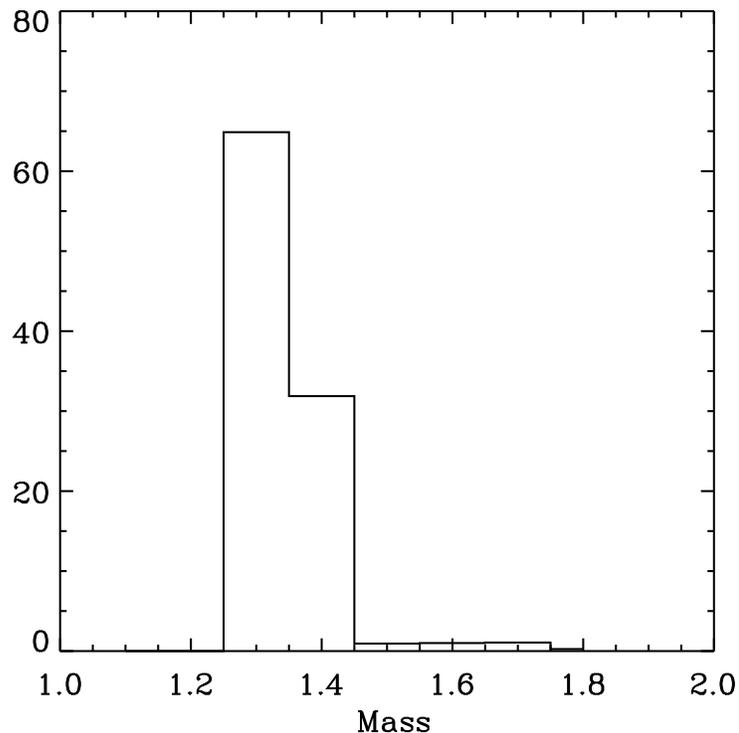}}}
\caption{Mass distribution for young close-by NSs. Stars were distributed
in eight bins from 1.1 to 1.8 solar masses. The vertical axis shows
percentage in each bin.
}

\end{figure}

This last point requires some more comments. 
Performing a population synthesis of cooling NSs demands for
the NS mass spectrum, since cooling curves depend on mass (e.g.
\opencite{kam02}). As noted by \inlinecite{whw02}, at present
models
do not allow a precise determination of the NS mass spectrum.
However, given the dependence of the cooling curves on mass (see below), 
even a rough estimate is enough for the case at hand. In our calculations
we use cooling curves for NS masses in the range
$1.1M_{\odot}< M< 1.8M_{\odot}$; masses are grouped in eight bins. 
Cooling models show that there
is a critical value for the mass ($\sim 1.35 M_{\odot}$ in
the case of \inlinecite{kam02}, the exact value
depends on model assumptions) across which
the cooling history significantly changes. 
NSs with masses below the critical value have similar cooling histories
and remain hot for a relatively long time ($T=10^5$~K after 4.25 Myrs). 
Intermediate mass stars ($\sim 1.4$~--~$1.5\, M_{\odot}$) cool down to
$10^5$~K in about the same time 
but have lower temperatures during the first million years
of their evolution
in comparison with less massive stars.
NSs with masses $M>1.5\, M_{\odot}$ experience much faster cooling and
become completely invisible (at X-ray energies) in a few hundred thousand
years or even earlier. 

\begin{figure}
\epsfxsize=\hsize
\centerline{{\epsfbox{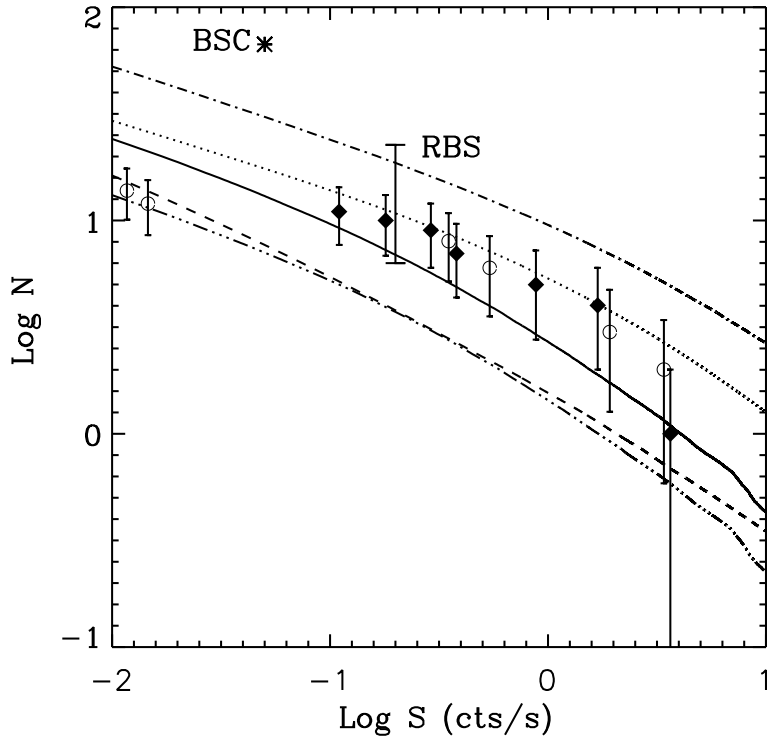}}}
\caption{Relative contribution of the "atmospheric effect"
and mass spectrum are shown. Dotted line --- the "old" model for 
disc+Belt contribution. Dashed line --- the "old" model without a Gould
Belt. Dot-dot-dashed (the lowest line)
--- "atmospheric effect". Dot-dashed --- effect of the new mass
spectrum. Solid line --- both effects together. 
Symbols which show observational points (filled diamonds or open circles)
are in correspondence with the type of the faintest object 
(a ROSAT  INS or not)
which contributes to the total number at the specified count rate.
RBS and BSC are limits on the number of bright INSs in ROSAT data
obtained by Schwope et al. (1999) and Rutledge et al. (2003) respectively.
}
\end{figure}

To our end, the most important point is estimating the number of NSs
above and below the critical mass. In our discrete description this
amounts to assess the number of stars in the first three mass bins
relative to remaining five. In order to do so, we proceed as follows. 
At first we take massive stars closer than 500 pc (i.e. with known
parallax $>$ 0.''002) from the Hipparcos catalog \cite{esa}.
Stars from B2 to O8 are considered here to be NS progenitors.
Spectral classes presented in the catalog are then transformed into
masses,
although we are aware that this is a very rough procedure.
NS masses are finally obtained from the  model described in
\inlinecite{tim96} and \inlinecite{whw02}. 
To do it we use a fit of fig.~14 in \inlinecite{whw02} and assumed
that all stars less massive than $\sim 11\, M_{\odot}$ produce
NSs of the same mass, i.e. 1.27~$M_{\odot}$. In all other cases 
the baryonic NS mass is calculated from the mass
of the progenitor according to 

\begin{equation}
M_{bar}=\cases{0.067M+0.567 & $11M_\odot < M < 15 M_\odot$\cr
& \cr
{\rm const}=1.567 & $15M_\odot \leq M \leq 20 M_\odot$\cr
& \cr
0.0867M-0.167 & $M > 20 M_\odot$\, .\cr}
\end{equation}
The NS gravitational mass (which is used in our calculation) is
calculated according to $M_{bar}-M_{grav}=0.075M^2_{grav}$
(\opencite{tim96}), here and in the formula below
masses are in the solar units.
All stars from the solar proximity
contributed to the final distribution with some coefficient, inversely
proportional to their lifetime ($\log t=9.9-3.8\log M+\log^2 M$).

Within the 500 pc sphere the number of progenitors with $M <
13.85\, M_\odot$ (which give a $1.35\, M_\odot$ NS) is about twice
higher than expected from the Salpeter mass function. Such an
enhancement is mostly connected with the Gould Belt. 
We find that about 2/3 of NSs have mass $< 1.35M_\odot$ and that most NSs
fall into the 1.3 and $1.4M_\odot$ bins (see fig.~2). The contribution of
massive ($>1.5\, M_\odot$) NSs is negligible (about 3\% by number).
This is a special feature of the solar proximity.

This mass spectrum is
in reasonable correspondence with mass determinations in binary
radio and X-ray
pulsars. We note, that the peak at 1.3  $M_\odot$ is due to the assumption
(see \opencite{tim96})
that all NSs below $\sim 11\, M_\odot$ produce NSs of
nearly the same mass, $\sim 1.27\, M_\odot$. However, smearing the
peak over the first three mass bins would produce about the same 
results for Log~N~--~Log~S since the cooling curves for these masses are
very similar in the time interval of interest for our calculations
(see \opencite{kam02}).
Although it represents just a rough estimate, this spectrum is better, in
our opinion, than
the flat one we used before, being closer to the one
obtained
from the  mass determinations in binary systems (mostly in binary radio
pulsars),
and in general it is in better correspondence with expectations about young
NSs in the solar vicinity.

\begin{figure}
\epsfxsize=\hsize
\centerline{{\epsfbox{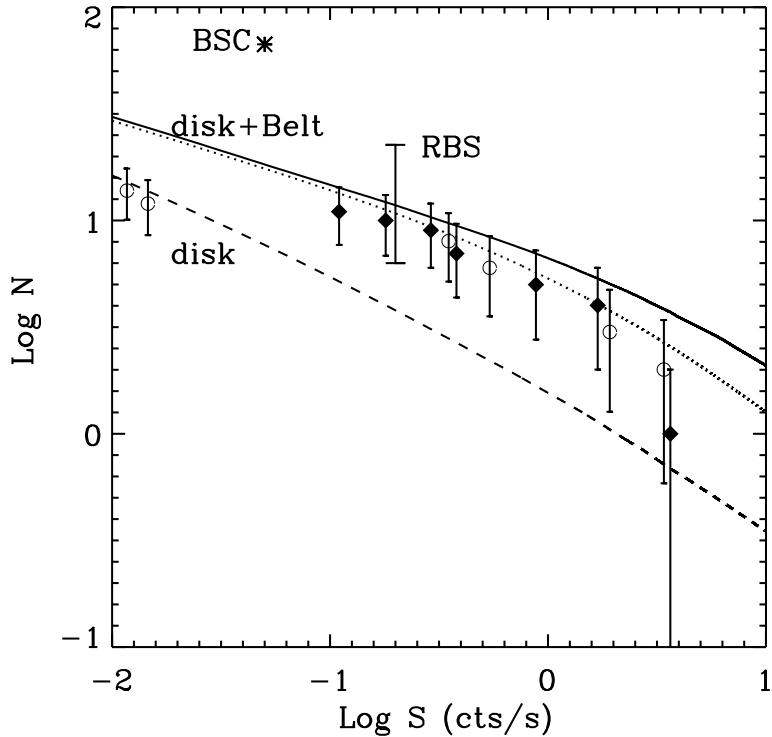}}}
\caption{Contributions of variations of initial spatial distribution and
kick velocity distribution are shown. Solid line: $R_{Belt}=300$ pc and
kick velocity distribution as in Arzoumanian  et al. (2002).
}
\end{figure}

\section{Results}

The Log~N~--~Log~S distributions
for young cooling INSs originated from the Gould Belt and circumsolar parts
of the galactic disc have been calculated for different parameters 
characterizing the NSs mass, velocity, and spatial distributions.  
The final variant, shown in fig.~5, includes all the four
modifications discussed in the previous section.
For comparison observational points representing 
the Log~N~--~Log~S distribution of isolated close-by 
NSs are also shown. These sources include
the seven ROSAT INSs ("the Magnificent seven"), several 
young close-by radio pulsars with detected thermal radiation 
("the three musketeers" and PSR B1929+10)
and Geminga together with the Geminga-like object 3EG~J1835+5918
(see Paper I for details).
Symbols for the observed data are in correspondence with the type
of the faintest object which contributes at a given flux (filled: ROSAT
INSs, empty: other sources). Error bars represent poissonian errors.
The two limits on the number of INSs in the ROSAT data obtained
by \inlinecite{rut03} (labeled BSC) and \inlinecite{rbs99} (labeled RBS)
are also shown.

\begin{figure}
\epsfxsize=\hsize
\centerline{{\epsfbox{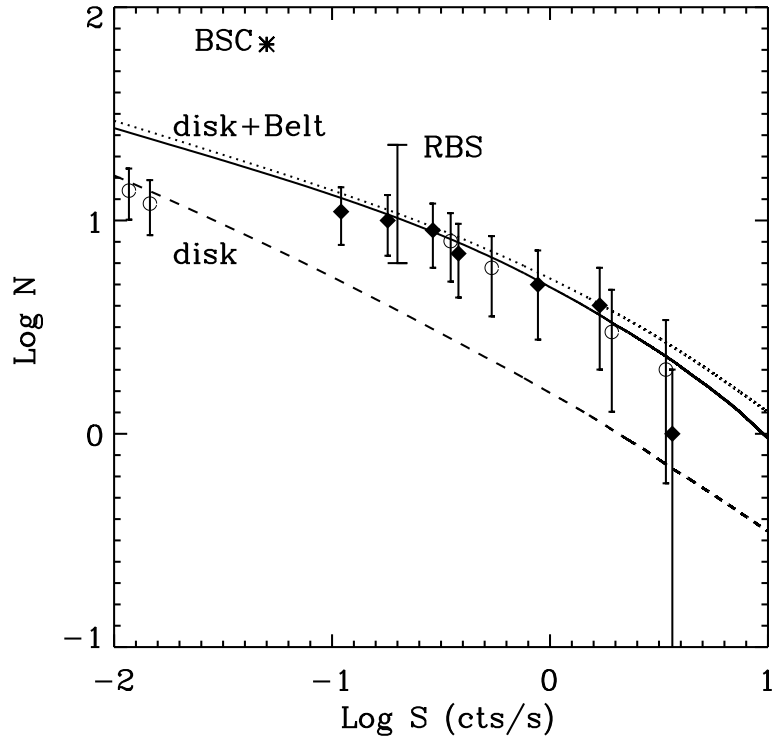}}}
\caption{A "new" model (solid line: $R_{Belt}=300$ pc, new mass spectrum,
an "atmospheric effect", new kick velocity distribution) in comparison
with older results (dotted line --- disc+Belt, dashed line --- only disc).
}

\end{figure}

Let us discuss first the relative effect of each of the 
four modifications
mentioned above. Clearly, a reduced surface emissivity and higher average
kick
velocity act in decreasing the number of observable sources at a given flux, 
while a higher fraction of
low mass (i.e. hotter) NSs and a more compact initial distribution
tend to increase it.

In fig.~3 we compare our previous result for the disc alone and
disc-plus-Belt with the new ones for disc-plus-Belt. To obtain these
new curves we considered either a smaller radiation radius
and flat mass spectrum  (dot-dot-dashed
curve) or the new mass spectrum together with standard
emissivity (dot-dashed curve).
One can see that the reduced emissivity and the new peaked
mass
spectrum move the Log~N--~Log~S curve (down and up, respectively) by
nearly half order of magnitude each,
with a net combined effect of slightly decreasing 
the number of observable sources. Here the spatial distribution of
NS progenitors and kick velocities are the same as in the original 
calculations. 
From the next figure (fig.~4) it is apparent that the same
effect is produced when a more compact
initial
distribution and
higher kick velocities are introduced, keeping the original assumptions 
about the star emissivity and mass spectrum.
Together these two modifications tend to slightly increase the number of
observable sources.

Our final results for Log~N~--~Log~S are presented in fig.~5.
Here all four effects are taken into account.
The general conclusion is that modifications do not change our
results significantly.
Our estimate is well below the BSC limit by \inlinecite{rut03}
and in correspondence with the RBS limit \cite{rbs99}.

\begin{figure}
\epsfxsize=\hsize	  
\centerline{{\epsfbox{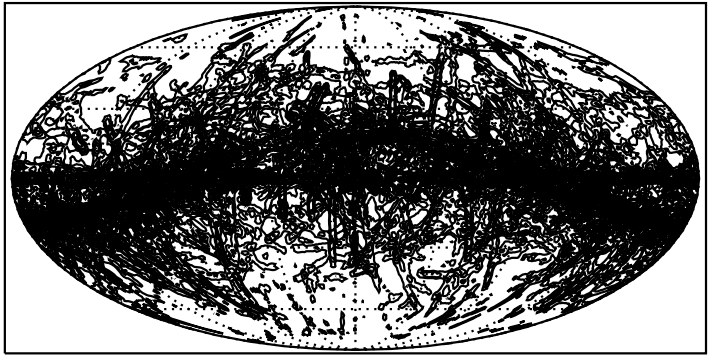}}}
\caption{Projected distribution of cooling INSs in the sky in galactic
coordinates. Only sources with count rate $>0.05$ cts~s$^{-1}$ are
accounted for. The total number of such sources is $\sim 17$ (see fig.~5).
The plot shows contours of constant INS number density per square
degree. Darker areas close to the Belt or/and to the
galactic plane correspond to $\sim 0.001$ sources/square degree. 
The presence of the Belt produces a tilt
in the higher projected density region which is visible in the figure. 
}

\end{figure}

\section{Discussion}

In this section we briefly discuss the spatial distribution of observable
INSs. Despite the main focus of our work has been of the
Log~N~--~Log~S curve, the present distribution of cooling INSs on the
sky is obtained as a by-product of our
evolutionary code. For illustrative purposes we report in fig.~6 the
projected spatial distribution of relatively bright coolers (count rate 
$>0.05$ PSPC cts~s$^{-1}$) in galactic coordinates.
This result should be taken with care as far as we used a simplified
initial
spatial distribution for the progenitors and did not take into account
any detailed ISM structure around the Sun.
The figure just illustrates some general features of the distribution.
The two main features of our model are appearent from the plot of
fig.~6 which shows that the highest projected density of sources
is close to the galactic plane. The presence of the Belt
produces the tilt of the high density region towards low galactic
latitudes.
All small scale details are due to individual tracks of calculated INSs 
and should not be treated as predictive of any ``fine structure''.

\begin{figure}
\epsfxsize=\hsize	  
\centerline{{\epsfbox{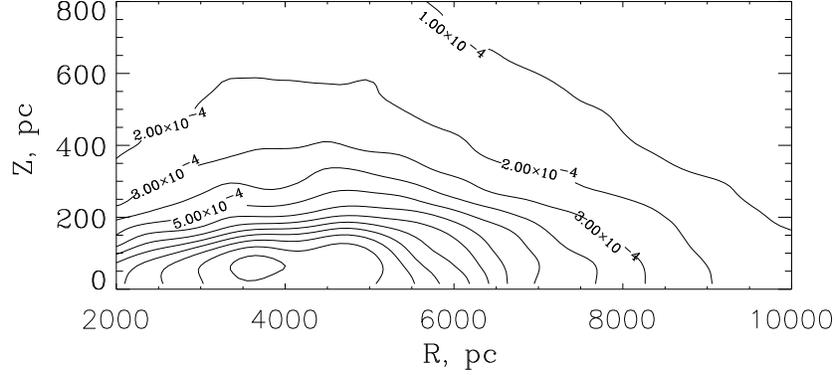}}}
\caption{Distribution of all isolated NSs in the Galaxy in the $R-z$ plane.
The data is calculated by a Monte Carlo of $>$~10000 individual tracks
on a fine grid (10 pc in $z$ direction and 100 pc in $R$ direction).
Curves were smoothed,
all irregularities are of statistical nature.
Kick velocity is assumed following Arzoumanian  et al. (2002).
NSs are born in the thin disc with semithickness 75 pc.
No NS born inside $R=2$~kpc and outside $R=16$~kpc are taken into account.
NS formation rate is assumed to be proportional to the square of the ISM
density at the birthplace. Results are normalized to have in total 5~$10^8$
NSs born in the described region.  Density contours are shown with a step 
$0.0001$~pc$^{-3}$.
}
\end{figure}

\begin{figure}
\epsfxsize=\hsize	  
\centerline{{\epsfbox{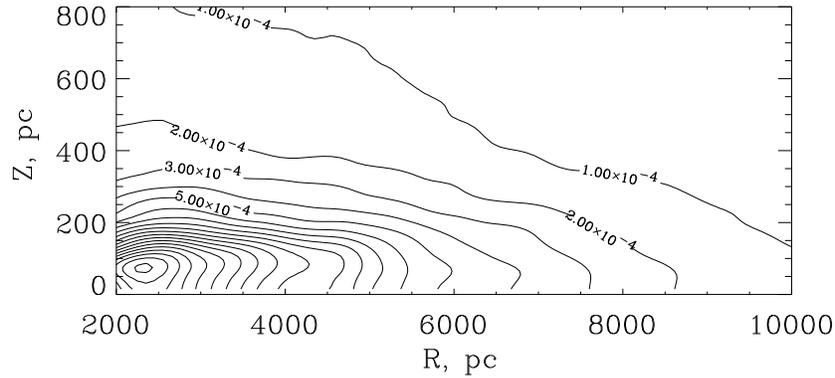}}}
\caption{Distribution of all isolated NSs in the Galaxy in the $R-z$ plane.
All parameters are as in the previous figure except the distribution of NS
formation rate, it is asusmed to be proportional to 
[${\rm exp}(-z/75\, {\rm pc})\, {\rm exp}(-R/4\, {\rm kpc})$].
Curves were smoothed as in the previous picture.
Is it clearly seen that in that case NSs are stronger concentrated towards
the galactic center, then in the case of NS formation rate proportional to
the square of the ISM density. 
}

\end{figure}

As expected, sources are strongly concentrated towards the galactic plane
and the Gould
Belt. Only about 12\% of sources with ROSAT count rate $>0.01$
cts~s$^{-1}$ are found at $\vert b\vert > 40^{\circ}$.
About 20\% of sources
lie outside the belt $\pm 30^{\circ}$ from the galactic plane, while
$\sim 50$\%  are expected to be within $\pm 12^{\circ}$ from the
plane of the Galaxy (brighter sources are more strongly concentrated
towards the
galactic plane and the Gould Belt since they correspond to younger INSs).
Although the very strong concentration towards the galactic plane may
reflect the assumption that NSs are born exactly in the (infinitesimally
thin) galactic disc, the source distribution at higher latitudes 
should be real.

Finally we would like to stress that the distribution of young NSs around
the Sun is definitely different
from the full NS spatial distribution, which is dominated by old stars
(age $> 10^7$~yrs). The latter is shown in figs.~7 and 8 for two different
assumptions about NS formation rate distribution.

We do not expect new identifications of bright ($>0.1$ cts~s$^{-1}$)
sources at large galactic latitudes. Most of the unidentified objects
(still there should be tens of them for count rates
$>0.01$ ROSAT cts~s$^{-1}$) should be in
crowded fields at $\pm 30^\circ$ from the plane of  the Milky Way.
Some can be identified as EGRET sources.

\acknowledgements

We want to thank Dmitry Yakovlev and his colleagues
for putting their cooling model to our disposal.
SP thanks Isabelle Grenier for discussions.
The work of MP and SP was partly supported by
the RFBR grant 03-02-16068 and by the program "Universities of Russia"
grant  02.03.013/2.
The work of MC, SP, AT and RT was partially supported by the
Italian Ministry for Education, University and Research (MIUR) under grant
COFIN-2000-MM02C71842.

\end{article}
\end{document}